\documentstyle[12pt]{article}

\begin{document}
\def\bea{\begin{eqnarray}}
\def\eea{\end{eqnarray}}
\title{\bf {Casimir stress for cylindrical shell in de-Sitter space }}
\author{
M.R. Setare  \footnote{E-mail: rezakord@yahoo.com}
  \\
{ Department of Science, Physics group, Kordestan University,
Sanandeg,  Iran  }\\ and \\ {Institute for Theoretical Physics and
Mathematics, Tehran, Iran}\\and \\{ Department of Physics, Sharif
University of Technology, Tehran, Iran }}
%\date{\small{\today}}
\maketitle
\begin{abstract}
The Casimir stress on cylinderical shell  in de Sitter background
for massless scalar field satisfying Dirichlet boundary conditions
on the cylinder is calculated. The metric is written in
conformally flat form to make maximum use of the Minkowski space
calculations. Different cosmological constants are assumed for the
space inside and outside the cylinder to have general results
applicable to the case of cylindrical domain wall formations in
the early universe.
 \end{abstract}
% \begin{document}
\newpage
% \vspace*{10mm}

 \section{Introduction}
The Casimir effect is one of the most interesting manifestations
  of nontrivial properties of the vacuum state in quantum field
  theory [1,2]. Since its first prediction by
  Casimir in 1948 \cite{Casimir} this effect has been investigated for
  different fields having different boundary geometries [4-7]. The
  Casimir effect can be viewed as the polarization of
  vacuum by boundary conditions or geometry. Therefore, vacuum
  polarization induced by a gravitational field is also considered as
  Casimir effect. The types of boundary and conditions that have been most often
  studied are those associated to well known problems, e.g. plates, spheres, and
  vanishing conditions, perfectly counducting conditions, etc. The cylindrical problem with
  perfectly counducting conditions was first considered in \cite{Mil}, for a recent study
  ref.\cite{{Gos},{Mil1}}.   \\
  In the context of hot big bang cosmology, the unified theories
  of the fundamental interactions predict that the universe passes
  through a sequence of phase transitions. These phase transitions
  can give rise to string structures determined by the topology of
  the manifold $M$ of degenerate vacuua \cite{{zel},{kib},{viel}}.
  If $M$ is disconnected, i.e. if $\pi(M)$ is nontrivial, then one
  can pass from one ordered phase to the other only by going
  through a domain wall. If $M$ has two connected components, e.g.
  if there is only a discrete reflection symmetry with
  $\pi_{0}(M)=Z_{2}$, then there will be just two ordered phase
  separated by a domain wall.\\
  The time evolution of topological defects have played an
  important role in many branches of physics, e.g., vortices in
  superconductors \cite{hu} and in superfluid \cite{do}, defects
  in liquid crystals \cite{ch}, domain wall \cite{{ar1},{ar2}},
  cosmic string \cite{{kib},{viel}} and a flux tube in QCD
  \cite{ba}.\\
   Casimir effect in curved space-time has not been studied extensively.
  Casimir effect in the presence of a general relativistic domain
  wall is considered in \cite{set} and a study of the relation
  between trace anomaly and the Casimir effect can be found in
  \cite{set1}. Casimir effect may  have interesting implications for the early
  universe. It has been shown, e.g., in \cite{Ant} that a closed Robertson-Walker
  space-time in which the only contribution to the  stress tensor comes from Casimir energy
   of a scalar field is excluded. In inflationary models, where the
   dynamics of bubbles may play a major role, this dynamical
   Casimir effect has not yet been taken into account. Let us
   mention that in \cite{set2} we have investigated the
   Casimir effect of a massless scalar field with Dirichlet
   boundary condition in spherical shell having different vacuua
   inside and outside which represents a bubble in early universe with
   false/true vacuum inside/outside. In this reference the sphere
   have zero thickness. In another paper \cite{set3} we have extended the
   analysis to the spherical shell with nonvanishing thickness.
    Parallel plates  immersed in different de Sitter
  spaces in- and out-side is calculated in \cite{set4}.  \\
  Our aim is to calculate the Casimir stress on a cylindrical shell in which the coordinate
  $z$ is absent with
  constant comoving radius having different vacuua inside and outside, i.e. with
   false/true vacuum  inside/outside. Our model may be used to study the effect of
   the Casimir force on the dynamics of the cylindrical domain wall appearing in
  the simplest Goldston model. In this model potential of the scalar field
  has two equal minima corresponding to degenerate vacuua. Therefore, scalar field
  maps points at spatial infinity in physical space nontrivially into the
  vacuum manifold \cite{vil1}. Domain wall structure occur at the boundary between
  these regions of space. One may assume that the outer regions of cylinder are
  in $\Lambda_{out}$ vacuum corresponding to degenerate vacuua in domain wall configuration.
   In section two we calculate the stress on
  cylinder with Dirichlet  boundary conditions. The case of
  different de Sitter vacuua inside and outside the cylinder, is
  considered in section three. The last section concludes and
  summarizes the results.

\section{Scalar Casimir effect for a cylindrical shell in flat space }
 In this section we calculate the Casimir energy of a
massless scalar field in flat space which
satisfies Dirichlet boundary condition on a cylinder.\\
The Casimir energy of a scalar field is given by
\begin{equation}
E_{0}=\frac{1}{2}\zeta(s-1/2)\mu^{2s},
  \end{equation}
where
\begin{equation}
\zeta(s)=\sum_{n}\lambda_{n}^{-s},
  \end{equation}
is the zeta function of the corresponding Laplace operator. The
arbitrary parameter $\mu$ has the dimension of a mass.\\
In our problem one can think of the circular problem in the plane,
because in the problem which we consider here, the coordinate $z$
is absent. The zeta function for interior region of circle is
given by \cite{les}
 \begin{equation}
\zeta^{int}(s)=\frac{1}{a}[-\frac{16+\pi}{128
\pi}(\frac{1}{s+1}+\ln a)+\frac{1}{8\pi}\frac{1}{s+1}+0.00985],
  \end{equation}
and for exterior region we have
\begin{equation}
\zeta^{out}(s)=\frac{1}{a}[\frac{16-\pi}{128
\pi}(\frac{1}{s+1}+\ln a)-\frac{1}{8\pi}\frac{1}{s+1}-0.0085].
  \end{equation}
The eigenvalues $\lambda_{n}$ which enter the zeta function (2)
are determined by
\begin{equation}
\nabla \varphi_{n}(x)=\lambda_{n}\varphi_{n}(x).
\end{equation}
Therefore, using Eqs.(1,2), the Casimir energies inside and
outside of cylinder are give by
 \begin{equation}
E_{in}=\frac{1}{2\mu} \zeta^{int}(-1)=\frac{1}{2a \mu
}[-\frac{16+\pi}{128 \pi}(\frac{1}{\varepsilon}+\ln
a)+\frac{1}{8\pi}\frac{1}{\varepsilon}+0.00985],
  \end{equation}
 \begin{equation}
E_{out}=\frac{1}{2\mu} \zeta^{out}(-1)=\frac{1}{2a \mu
}[\frac{16-\pi}{128 \pi}(\frac{1}{\varepsilon}+\ln
a)-\frac{1}{8\pi}\frac{1}{\varepsilon}-0.0085].
  \end{equation}
Each of the energies for inside and outside of the cylinder is
infinite, and cutoff dependent. The Casimir energy $E$ is the sum
of Casimir energies $E_{in}$ and $E_{out}$ for inside and outside
of the cylinder.
 \begin{equation}
E=E_{in}+E_{out}=\frac{1}{2a \mu}[ \frac{-1}{64}(
\frac{1}{\varepsilon}+\ln a) +0.00135].
\end{equation}
As one can see, the Casimir energy $E$ is also infinite and
dependent to the cutoff $\varepsilon$. At this stage we introduce
the classical energy for inside and outside separately and try to
absorb divergent parts into these classical energies. This
technique of absorbing an infinite quantity into a renormalized
physical quantity is familiar in quantum field theory and quantum
field theory in curved space \cite{davies}. Here we use a
procedure similar to that of bag model \cite{{bord3},{bord4}} (
to see application of this renormalization procedure in Casimir
effect problem in curved space refer to \cite{{set2},{set3}}). The
classical energy of cylinder in which the $z$ coordinate is absent
may be written as
\begin{equation}
E_{class}=\sigma a^{2}+Fa+K+\frac{h}{a}.
\end{equation}
The total energy of the cylinder inside and outside may be written
as
 \begin{equation}
\tilde{E}_{in}=E_{in}+E^{in}_{class}
\end{equation}
\begin{equation}
\tilde{E}_{out}=E_{out}+E^{out}_{class}.
\end{equation}
In order to obtain a well defined result for the Casimir energies
inside and outside cylinder, we have to renormalize the parameter
$h$ of classical energy according to below
\begin{equation}
h^{in}\rightarrow h^{in}+\frac{1}{256 \mu \varepsilon}
\end{equation}
\begin{equation}
h^{out}\rightarrow h^{out}+\frac{1}{256 \mu \varepsilon}.
\end{equation}
Hence the effect of the vacuum fluctuation of scalar quantum field
is to change, or renormalize parameter $h$ of classical energy
inside and outside of cylinder. Therefore, we rewrite Eqs.(10,11)
as
 \begin{equation}
\tilde{E}_{in}=E_{in}+\tilde{E}^{in}_{class}=\frac{1}{2a\mu}[0.00985-
\ln a( \frac{1}{128}+\frac{1}{8\pi})]+\frac{h^{in}}{a},
\end{equation}
\begin{equation}
\tilde{E}_{out}=E_{out}+\tilde{E}^{out}_{class}=\frac{1}{2a\mu}[-0.0085+
\ln a( \frac{-1}{128}+\frac{1}{8\pi})]+\frac{h^{out}}{a}.
\end{equation}
We finally obtain for the total zero point energy of our system
 \begin{equation}
\tilde{E}=\frac{1}{2a \mu}[0.00135-\frac{\ln a}{64}].
\end{equation}
Once, the infinite terms have been removed from $E$ in Eq.(8), the
remainder is finite and will be called the renormalized Casimir
energy. The Casimir stress on the cylinder due to the boundary
conditions is then obtained
\begin{equation}
\frac{\bar{F}}{A}=\frac{-1}{2\pi a}\frac{\partial
\tilde{E}}{\partial a}=\frac{1}{4\pi \mu a^{3}}[(
0.00135-\frac{\ln a}{64}) +\frac{1}{64 }].
\end{equation}
\section{Cylindrical shell in de Sitter space}
Consider now a massless scalar filed in de Sitter space-time which
satisfies Dirichlet boundary condition on a cylindrical shell. To
make the maximum use of the flat space calculation we use the de
Sitter metric in conformally flat form
\begin{equation}
ds^{2}=\frac{\alpha^{2}}{\eta^{2}}[d\eta^{2}-\sum_{i=1}^{3}(dx^{i})
^{2}],
\end{equation}
where $\eta$ is the conformal time
\begin{equation}
-\infty <\eta < 0.
\end{equation}
The constant $\alpha$ is related to the cosmological constant as
\begin{equation}
\alpha^{2}=\frac{3}{\Lambda}.
\end{equation}
Now we consider the pure effect of vacuum polarization due to the
gravitational field without any boundary conditions (to see such
problem for spherical shell and parallel plate geometry refer to
\cite{{set2},{set3},{set4}}). The renormalized stress tensor for
massless scalar field in de Sitter space is given by
\cite{{davies},{dow}}
\begin{equation}
< T^{\nu}_{\mu}>=\frac{1}{960
\pi^{2}\alpha^{4}}\delta^{\nu}_{\mu}.
\end{equation}
The corresponding effective pressure is
\begin{equation}
P=-< T^{1}_{1}>=-< T^{r}_{r}>=-\frac{1}{960 \pi^{2}\alpha^{4}},
\end{equation}
valid for both inside and outside the cylinder. Hence the
effective force on the cylinder due to the gravitational vacuum polarization is zero. \\
Now, assume there are different vacuua inside and outside
corresponding to $\alpha_{in}$ and $\alpha_{out}$ for the metric
(18). Now, the effective pressure created by gravitational part
(22), is different for different part of space-time
\begin{equation}
P_{in}=-< T^{r}_{r}>_{in}=-\frac{1}{960
\pi^{2}\alpha^{4}_{in}}=\frac{-\Lambda_{in}^{2}}{8640 \pi^{2}},
\end{equation}
\begin{equation}
P_{out}=-< T^{r}_{r}>_{out}=-\frac{1}{960
\pi^{2}\alpha^{4}_{out}}=\frac{-\Lambda_{out}^{2}}{8640 \pi^{2}}.
\end{equation}
Therefore the gravitational pressure over shell, $P_{g}$, is given
by
\begin{equation}
P_{g}=P_{in}-P_{out}=\frac{-1}{8640 \pi^{2}}(
\Lambda_{in}^{2}-\Lambda_{out}^{2}).
\end{equation}
Now we consider the effective pressure due to the boundary
condition.  Under the conformal transformation in four dimensions
the scalar
  field $\Phi(x,\eta)$ is given by
    \begin{equation}
  \bar\Phi(x,\eta)=\Omega^{-1}(x,\eta)\Phi(x,\eta),
  \end{equation}
  with the conformal factor given by
  \begin{equation}
  \Omega(\eta)=\frac{\alpha}{\eta}.
  \end{equation}
 And assuming a canonical quantization of the scalar field, and using
  the creation and annihilation operators $a_{k}^{\dagger}$ and
  $a_{k}$, the scalar field $\Phi(x,\eta)$ is then given by
\begin{equation}
  \Phi(x,\eta)=\Omega(\eta)\sum_{k}[a_{k}
  \bar u_{k}(\eta,x)+a_{k}^{\dagger}\bar u_{k}^{\ast}(\eta,x)]
  \end{equation}
  The vacuum states associated with the modes $\bar u_{k}$ defined by
  $a_{k}|\bar 0\rangle=0 $, are called conformal vacuum. For the massless
  scalar field we are considering, the Green's function $\bar G$
  associated to the conformal vacuum $|\bar 0\rangle$ is given by the
  flat Feynman Green's function times a conformal factor \cite{{davies},{chit}}.
  The two-point Green's function, $G(x,t;x',t')$, is defined as the
  vacuum expectation value of the time-ordered product of two
  fields
  \begin{equation}
  G(x,t;x',t')=-\imath\langle 0|T \Phi(x,t)
 \Phi(x',t')|0 \rangle.
 \end{equation}
 Given the above flat space Green's function, we obtain
  \begin{equation}
 \bar G=-\imath\langle\bar{0}|T \bar \Phi(x,\eta)\bar
 \Phi(x',\eta^{'})|\bar{0}\rangle=\Omega^{-1}(\eta)\Omega^{-1}(\eta^{'})G.
 \end{equation}
Using this Green's function we can obtain the Casimir stress
inside and outside of the cylinder in de Sitter space
\begin{equation}
(\frac{\bar
F}{A})_{in}=\frac{\eta^{2}}{\alpha^{2}_{in}}(\frac{F}{A})_{in},
\end{equation}
\begin{equation}
(\frac{\bar
F}{A})_{out}=\frac{\eta^{2}}{\alpha^{2}_{out}}(\frac{F}{A})_{out}.
\end{equation}
Now, using Eqs.(14,15,17) and considering the zero point energy
inside and outside we can write
\begin{equation}
(\frac{\bar F}{A})_{in}=-\frac{1}{2\pi
a}\frac{\eta^{2}}{\alpha_{in}^{2}}\frac{\partial
\tilde{E}_{in}}{\partial a}= \frac{1}{4 \pi \mu a^{3}}\frac{\eta
^{2}}{\alpha_{in}^{2}}[ 0.00985+(1-\ln a) (
\frac{1}{128}+\frac{1}{8\pi})],
\end{equation}
\begin{equation}
(\frac{\bar F}{A})_{out}=-\frac{1}{2\pi
a}\frac{\eta^{2}}{\alpha_{out}^{2}}\frac{\partial
\tilde{E}_{out}}{\partial a}= \frac{1}{4 \pi \mu a^{3}}\frac{\eta
^{2}}{\alpha_{out}^{2}}[- 0.0085+(ln a-1) (
\frac{-1}{128}+\frac{1}{8\pi})].
\end{equation}
Therefore, the vacuum pressure due to the boundary condition
acting on the cylinder is given by \bea
 P_{b}&=&(\frac{\bar
F}{A})_{in}+(\frac{\bar F}{A})_{out}=\frac{ \eta^{2}}{4\pi \mu
a^{3}}[ \frac{1}{\alpha_{in}^{2}}[0.00985+( 1-\ln a) (
\frac{1}{128} + \frac{1}{8
\pi})]\\
&+&\frac{1}{\alpha_{out}^{2}}[-0.0085+(1-\ln a)
(\frac{1}{128}-\frac{1}{8\pi})]].\nonumber
 \eea
  The total pressure on the
circle, $P$, is then given by \bea
 P&=& P_{g}+P_{b}=\frac{1}{8640
\pi^{2}}( \Lambda_{out}^{2}-\Lambda_{in}^{2})+\frac{
\eta^{2}}{12\pi \mu a^{3}}( ( 1-\ln a)[\Lambda_{in} (
\frac{1}{128}+\frac{1}{8
\pi})\\
&+&
\Lambda_{out}(\frac{1}{128}-\frac{1}{8\pi})]+0.00985\Lambda_{in}-0.0085\Lambda_{out}).\nonumber
\eea The $\eta$- or time-dependence of the pressure is intuitively
clear due to the time dependence of the physical radius of
cylinder.
 Total pressure, may be negative or positive, depending on the
difference between the cosmological constant in the two parts of
space-time. Given a false vacuum inside of the cylinder  , and
true vacuum outside, i.e. $\Lambda_{in}> \Lambda_{out}$, if $1>\ln
a$, then the gravitational part is negative, and tends to contract
the cylinder. In contrast, the boundary part is positive and will
lead to the repulsive force. Therefore, the total effective
pressure on the cylinder may be negative, leading to a contraction
of the cylinder. The contraction, however, ends for a minimum of
radius of the cylinder, where both part of the total pressure are
equal. For the case of true vacuum inside the cylinder and false
vacuum outside, i.e $\Lambda_{in}< \Lambda_{out}$, the
gravitational pressure is positive. In this case, boundary part
can be negative or positive depending on the difference between
$\Lambda_{in}$ and $\Lambda_{out}$. Hence, the total pressure may
be either negative or positive.
\section{Conclusion}

 We have considered a cylinder in which the coordinate $z$ is absent in de Sitter background with a massless scalar field,
 coupled conformally to it, satisfying the Dirichlet
 boundary conditions with constant comoving radius. Our calculation
 shows that for the cylindrical shell with false vacuum inside and
 true vacuum outside, the gravitational pressure part is negative,
 but the boundary pressure part is positive. In contrast for the
 case of true vacuum inside the cylinder and false vacuum outside,
 the gravitational pressure is positive, and boundary part can be
 negative or positive depending on the difference between
 cosmological constant inside and outside of cylinder.
  The result may be of interest in the case of
 formation of the cosmic cylinderical domain walls in early
 universe.

\end{document}